\documentclass[sigconf]{acmart}

\usepackage[ruled]{algorithm2e}
\usepackage{stmaryrd}
\usepackage{graphicx}
\usepackage{color}
\usepackage{comment}
\usepackage{url}
\usepackage{wrapfig}
\usepackage{subfigure}
\usepackage{tabularx}
\usepackage{algpseudocode}
\usepackage{wrapfig}
\usepackage{balance}

\usepackage[T1]{fontenc}
\usepackage{lmodern}
\begin{document}

\acmYear{2018} 
\setcopyright{acmcopyright}
\acmConference[Packet Video'18]{23rd Packet Video Workshop}{June 12--15, 2018}{Amsterdam, Netherlands}
\acmPrice{15.00}
\acmDOI{10.1145/3210424.3210429}
\acmISBN{978-1-4503-5773-9/18/06}

\linespread{0.89}

\title{Dynamic Adaptive Point Cloud Streaming}

\author{Mohammad Hosseini}
\affiliation{\institution{University of Illinois at Urbana-Champaign\\ (UIUC)}}
\email{shossen2@illinois.edu}

\author{Christian Timmerer}
\affiliation{\institution{Alpen-Adria-Universit{\"a}t Klagenfurt, Bitmovin Inc.}}
\email{christian.timmerer@itec.uni-klu.ac.at}

\begin{abstract}
High-quality point clouds have recently gained interest as an emerging form of representing immersive 3D graphics. Unfortunately, these 3D media are bulky and severely bandwidth intensive, which makes it difficult for streaming to resource-limited and mobile devices. This has called researchers to propose efficient and adaptive approaches for streaming of high-quality point clouds.

In this paper, we run a pilot study towards dynamic adaptive point cloud streaming, and extend the concept of dynamic adaptive streaming over HTTP (DASH) towards DASH-PC, a dynamic adaptive bandwidth-efficient and view-aware point cloud streaming system. DASH-PC can tackle the huge bandwidth demands of dense point cloud streaming while at the same time can semantically link to human visual acuity to maintain high visual quality when needed. In order to describe the various quality representations, we propose multiple thinning approaches to spatially sub-sample point clouds in the 3D space, and design a DASH Media Presentation Description manifest specific for point cloud streaming. Our initial evaluations show that we can achieve significant bandwidth and performance improvement on dense point cloud streaming with minor negative quality impacts compared to the baseline scenario when no adaptations is applied.
\end{abstract}



\maketitle
\section{Introduction}
While traditional multimedia applications such as videos are still popular, there has been a substantial interest towards new media, such as VR and immersive 3D graphics. High-quality 3D point clouds have recently emerged as an advanced representation of immersive media, enabling new forms of interaction and communication with virtual worlds.

3D point clouds are a set of points represented in the 3D space, each associated with multiple attributes such as coordinate and color. They can be used to reconstruct a 3D object or a scene composing of various points. point clouds can be captured using camera arrays and depth sensors in various setups, and may be made up to billions of points in order to represent reconstructed objects in a high-quality manner.

Despite the promising nature of point clouds, these 3D media are highly resource intensive, and therefore are difficult to stream and render at acceptable quality levels. Therefore, a major challenge is how to efficiently transmit the bulky high-quality point clouds especially to bandwidth-constrained mobile devices. For raw dynamic point cloud frames, the data transmission rate for an application with 30 fps can be as high as 6 Gbps \cite{8i}. Therefore, there must be a balance between the requirements of streaming and the available resources. One of the challenges for achieving this balance is to meet this requirement without much negative impact on the user's viewing experience in an immersive environment.

While our work is motivated by the concepts presented by Dynamic Adaptive Streaming over HTTP (DASH) for adaptive video streaming, a semantic link between adaptive streaming of point clouds with user's view and limited bandwidth has not been fully developed yet for the purpose of bandwidth management and high-quality point cloud streaming. In this paper, we aim to utilize this semantic relation to study and extend the concept of adaptive streaming towards point cloud streaming. We propose DASH-PC, a dynamic adaptive view-aware and bandwidth-efficient point cloud streaming approach to tackle the high bandwidth requirements of dynamic point cloud models. To enable various quality representations, we sub-sample the dense point cloud frames spatially in the 3D space, and construct a point cloud-specific DASH-like manifest. Our preliminary study shows that we can achieve substantial improvement in streaming and rendering of dense point cloud data without noticeable quality impacts compared to the baseline scenario when no adaptations is applied.

The paper is organized as follows: in Section 2, we briefly cover some background and related work. In Section 3, we explain our methodology including the framework, sub-sampling, and visual acuity. Our experiments and evaluation results are presented in Section 4, while in Section 5 we conclude the paper and briefly discuss possible avenues for future work.

\section{Background and Related Work}

\subsection{Dynamic Adaptive Streaming}
One of the main approaches for bandwidth saving on bandwidth-intensive multimedia applications is adaptive streaming. Adaptive streaming is a process where the quality of a multimedia stream is altered in real-time while it is being sent from a server to a client. The adaptation may be the result of adjusting various network or device metrics. For example, with a decrease in network throughput, adaptation to a lower video bitrate may reduce re-buffering events and improve the user's experience.

Dynamic Adaptive Streaming over HTTP (DASH) specifically, also known as MPEG-DASH \cite{dash1}, is an ISO standard that enables client-driven adaptive video streaming whereby a client chooses a video segment with the appropriate quality (bit rate, resolution, etc.) based on its constrained resources such as bandwidth. The video content is stored on an HTTP server, and is accompanied by a Media Presentation Description (MPD) as a manifest of the available segments, bitrates, and other characteristics.

In this study, we extend the semantics of MPEG-DASH towards a dynamic point cloud streaming system that enables view-aware and bandwidth-aware adaptation through transmitting varied quality of point cloud data relative to the user's view, bandwidth, or other resources.

\subsection{View-Aware Media Streaming}
View-aware media adaptation has become a \textit{de facto} in adaptive multimedia streaming and media content prioritization, especially in the context of emerging media such as 360 videos, VR, and 3D media. In the context of 360 VR videos for example, various studies have been conducted to leverage view-aware adaptations to efficiently transmit, render, and display 360 videos to resource-limited VR headsets \cite{ism, ieeevr, timmerer, corbillon}. Similarly, the authors in \cite{mmsj} and \cite{mmve} studied view-aware streaming in the context of 3D tele-immersive systems. Their approach assigns higher quality to parts within users' viewport given the features of the human visual system.

In this paper, we follow similar concepts to propose view-aware adaptation techniques to reduce the bandwidth requirements of high-quality point cloud streaming.

\subsection{Point Cloud Compression}
While currently there is no work addressing dynamic and adaptive point cloud streaming, existing works are mostly centered around point cloud compression using geometry-based approaches. 

Google's Draco \cite{draco} project uses kd-tree data structure for quantifying and organizing points in the 3D space. Similarly, Schnabel and Klein in their work \cite{schnabel} proposed a solution based on an octree decomposition of space, where a point cloud is encoded in terms of octree cells. Overall, the two main important methods that recursively subdivide the bounding box are the kd-tree approach of Devillers and Gandoin \cite{devillers} and the octree-based approach by Huang \textit{et. al.} \cite{huang}. In the same domain, some methods have been proposed for the purpose of differential coding. In \cite{gumhold}, the authors propose a predictive approach to compress points in a spatially sequential order. Points are predicted in such a way that only corrective vectors are encoded. Similarly, in \cite{kammerl}, a modified octree data structure is used for differential encoding of spatial changes. On the MPEG side, a new ad-hoc group has been initiated for Point Cloud Compression, or in short, MPEG PCC \cite{mpegpcc}, aimed to cover topics in lossy and lossless point cloud compression.




\begin{figure}[!t]
\centering
\includegraphics[width=\columnwidth]{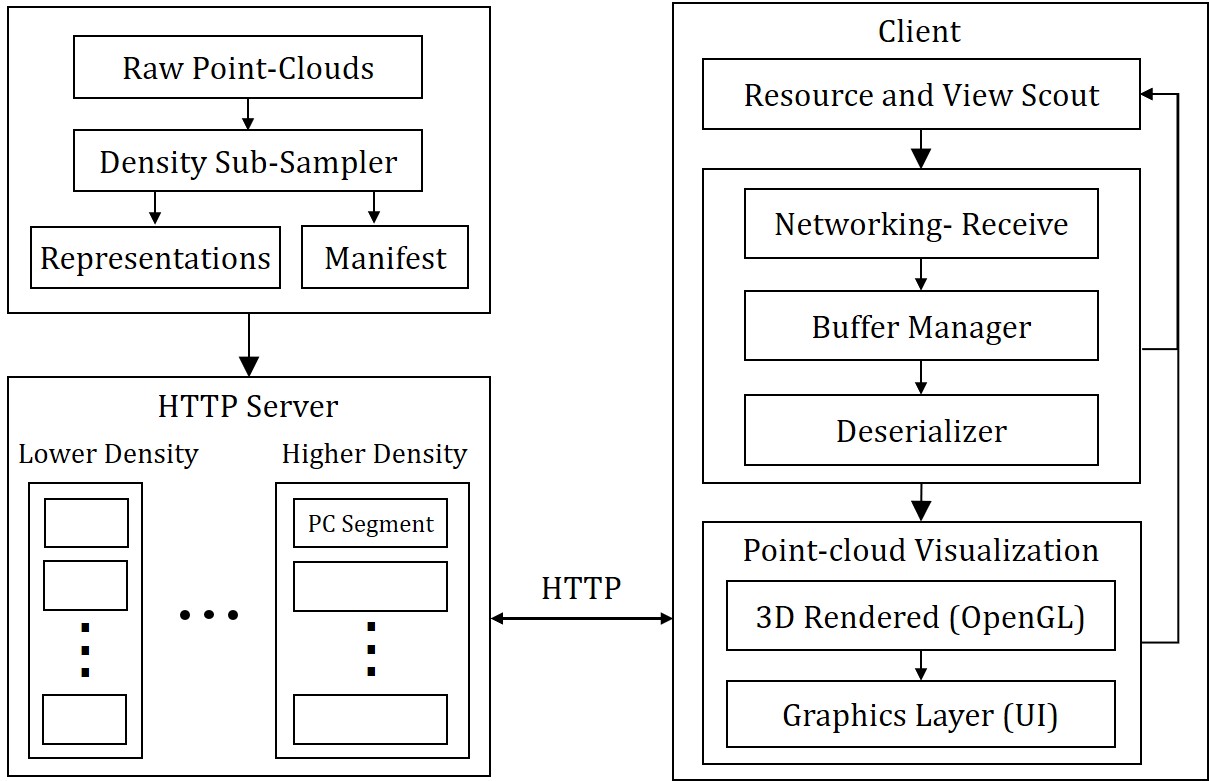}
\caption{DASH-PC architecture overview.}
\vspace{-0.2cm}
\label{framework}
\end{figure}

\begin{figure}[!t]
\centering
\includegraphics[width=\columnwidth]{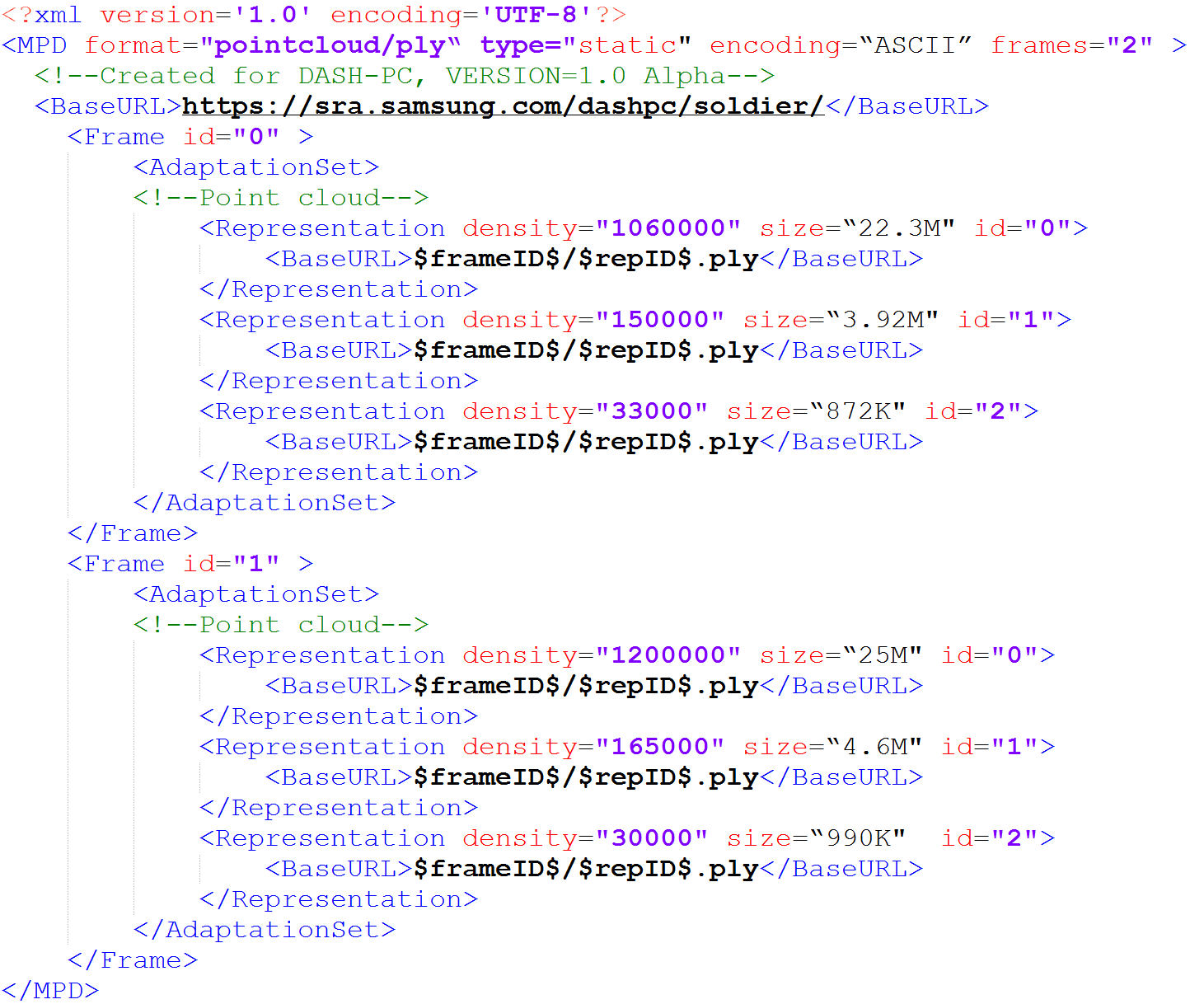}
\caption{An example DASH-PC manifest.}
\vspace{-0.5cm}
\label{mpd}
\end{figure}

\section{Methodology}
One of the major challenges in streaming dense point clouds is their high volume and therefore, high streaming bandwidth demands. To decrease the bandwidth requirements, one approach is to decrease the density of points in the 3D space. This not only leads to a smaller volume of points, but also decreases the GPU-assisted graphics rendering and the processing overhead. Decreasing the point cloud density however, might also cause reduction of visual quality depending on how the \textit{Level of Density} (LoD) is crucial in the context of the application. The aim of DASH-PC is to incorporate these features and allow the client to select the best LoD representation depending on its available resources, viewing preference, or any other application preferences. Density-based adaptation is particularly useful for the purpose of visual-compliant adaptations.

Our proposed framework can be used to efficiently transmit point clouds by allowing the client devices (or the server communicating with the client devices) to select point cloud frames with a specific specification, using standard DASH. A client can receive point cloud frames by receiving the segments through HTTP request/response communications, which allows to dynamically switch between different density representations. DASH-PC can provide a better quality through adaptation to varying network link conditions such as available bandwidth, or various device capabilities such as display resolution or processing power, as well as any user preferences such as the point cloud scale or user's view distance from the camera. Figure \ref{framework} illustrates an overview of the DASH-PC architecture.

Similar to DASH, the adaptation feature of DASH-PC is described by an XML-formatted manifest, which contains metadata required for the client to establish HTTP requests to the media server to retrieve point cloud models. Our designed DASH-PC manifest includes characteristics of the different point cloud representations stored on a standard HTTP server. The manifest is divided into separate frames, with each frame including a variety of adaptation sets, each providing information about multiple quality alternatives, also carrying information on the index of each frame, the frame's HTTP location, and LoD representations.

Figure \ref{mpd} illustrates an example DASH-PC manifest as used by our architecture. In the following, we briefly summarize the basic descriptors and attributes defined within the DASH-PC MPD.

\subsection{DASH-PC MPD Semantics}
For a better compliance with the semantics of MPEG-DASH, we re-use some of the DASH mime types in the DASH-PC manifest when possible. However, here we re-define them with the new point cloud concepts.
\subsubsection*{@format} specifies the point cloud container format (e.g., ply).

\subsubsection*{@encoding} specifies the point cloud encoding (ASCII or binary).

\subsubsection*{@frames} specifies the number of point cloud frames followed.

\subsubsection*{@type} shows if the manifest may be updated during a session ('static' (default) or 'dynamic').

\subsubsection*{@BaseURL} specifies the base HTTP URL of the point cloud.

\subsubsection*{Frame}
A frame typically represents a single point cloud model. For each frame, a consistent set of multiple adaptation sets of that point cloud model are available. The basic elements and attributes of a \textit{Frame} element are as follows:
\begin{itemize}
\item id: a unique identifier of the frame.
\item BaseURL: specifies the base HTTP URL of the frame.
\item AdaptationSet: specifies the sub-clause adaptation sets.
\end{itemize}
\subsubsection*{AdaptationSet}
Within a frame, a model is arranged into adaptation sets. An adaptation set consists of multiple interchangeable quality versions of a point cloud frame. For instance, adaptation sets can carry additional attributes to represent density sub-sampling methods, so that there may be one adaptation set for each density sub-sampling approach, data structure used, or compression techniques used to represent point clouds. Similarly, an adaptation set can carry additional attributes to account for a specified viewport, with each viewport having a separate adaptation set. The basic elements and attributes of an AdaptationSet element are as follows:
\begin{itemize}
\item id: a unique identifier of the adaptation set.
\item BaseURL: specifies the base HTTP URL of the adaptation set.
\item Representation: specifies the sub-clause representations.
\end{itemize}

\subsubsection*{Representation}
An adaptation set contains a set of representations, each describing a deliverable version of a point cloud frame model. A single point cloud representation is sufficient for rendering and visualization of a model. Typically, a client can switch from one representation to another at any time in order to adapt to its varying resources and parameters such as network bandwidth, energy budget, visual preferences, etc. A client can ignore representations with unsupported rendering or encoding specifications. For instance, if a client only renders binary point clouds, it can ignore ASCII-encoded representations. The basic elements and attributes of a \textit{Representation} element are as follows:
\begin{itemize}
\item id: unique identifier of the representation.
\item BaseURL: specifies the base HTTP URL of the representation.
\item density: specifies the density of the point cloud model in terms of the number of points.
\item size: specifies the size of the point cloud model in terms of the storage volume.
\item Segment: specifies the sub-clause segments.
\end{itemize}

\subsubsection*{Segment}
Each representation may contain one or more segments. Segments describe sub-models based on spatial segmentation which can be used to retrieve partial point cloud models. We use the notion of segments to better comply with the features of viewport adaptations. For instance, a set of specific segments might be visible for a specific viewport while they are not visible from another viewport. Therefore, additional attributes such as viewpoint position and orientation information might be accompanied. The basic elements and attributes of a \textit{Segment} element are as follows:
\begin{itemize}
\item id: unique identifier of the adaptation set.
\item BaseURL: specifies the HTTP URL of the segment.
\item density: specifies the density of the point cloud model in terms of the number of points.
\item size: specifies the size of the point cloud model in terms of the storage volume.
\end{itemize}

\subsection{Point Cloud Sub-sampling}

\begin{algorithm}[!t]
\begin{algorithmic}
\State $R$: the sub-sampling ratio
\State $\{P\}$: set of points in the original model
\State $|P|$: current number of points
\State $\{P'\}$: set of points in the sub-sampled model
\State $|P'| \gets |P| / R $: calculate the total number of points in the sub-sampled point cloud
\State $\forall p_i \in \{P\}: 0\leq i < |P|$:
\State \hspace{.5cm}${P} \gets Sort(\{P\}, X)$
\State \hspace{.5cm}${P} \gets Sort(\{P\}, Y)$
\State \hspace{.5cm}${P} \gets Sort(\{P\}, Z)$
\State $\forall j: 0\leq j < |P'|$:
\State \hspace{.5cm}${P'} \gets p_{(i\times R)}$
\State store the sub-sampled points in a new \texttt{.ply} container
\end{algorithmic}
 \caption{The first density sub-sampling process.}
 \label{alg1}
\end{algorithm}

\begin{figure}[!t]
    \vspace{-.3cm}
    \centering
    \includegraphics[width=.32\columnwidth]{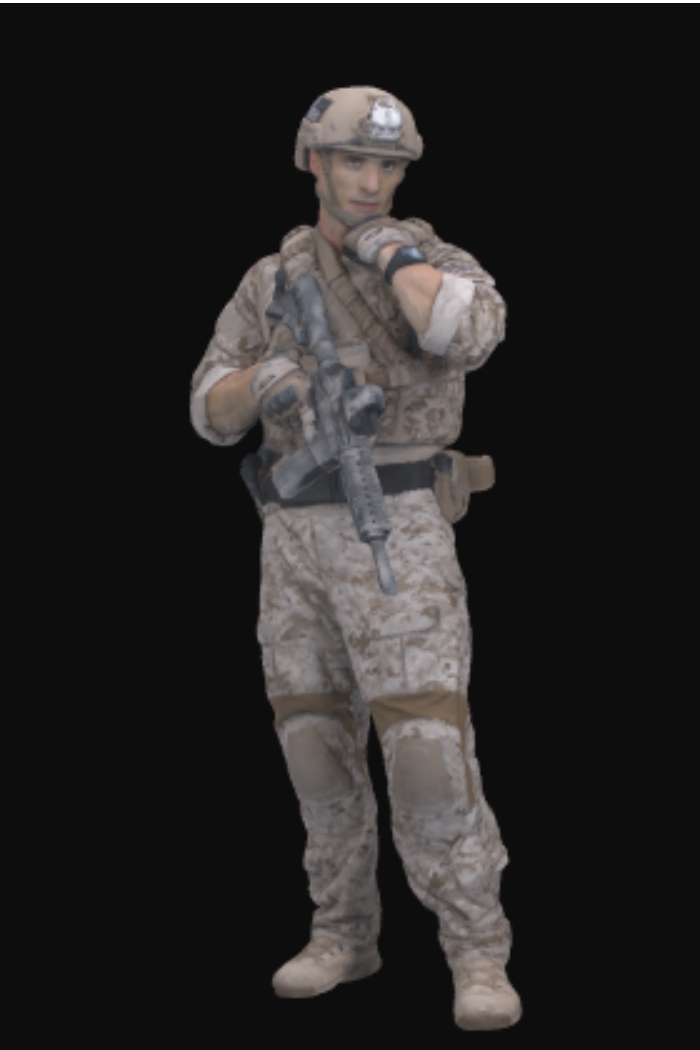}
    \includegraphics[width=.32\columnwidth]{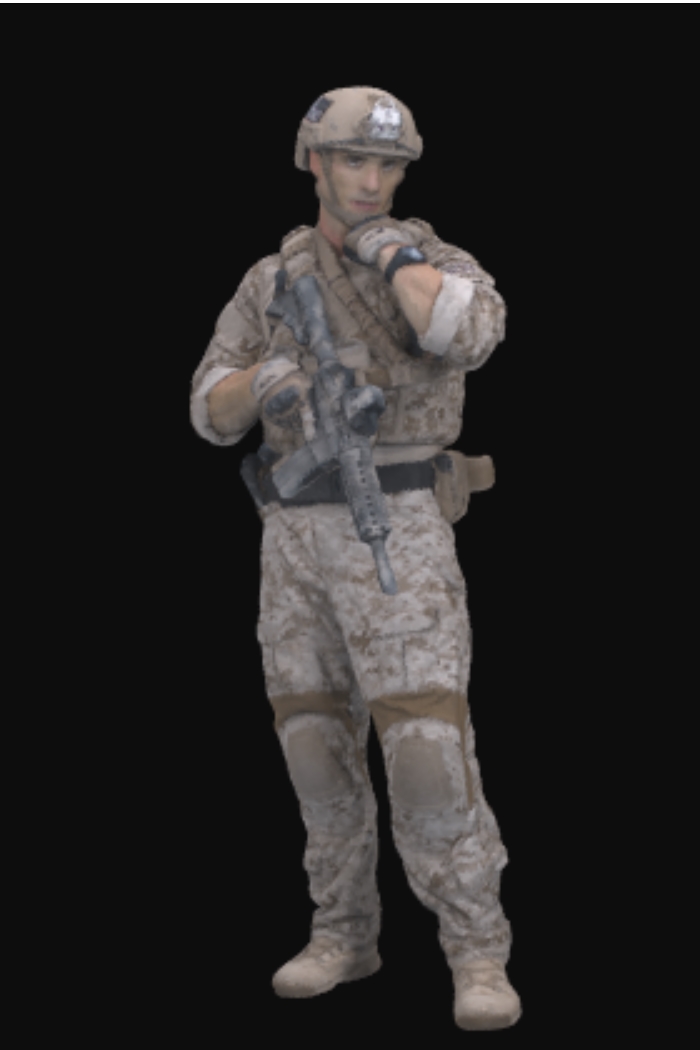}
    \includegraphics[width=.32\columnwidth]{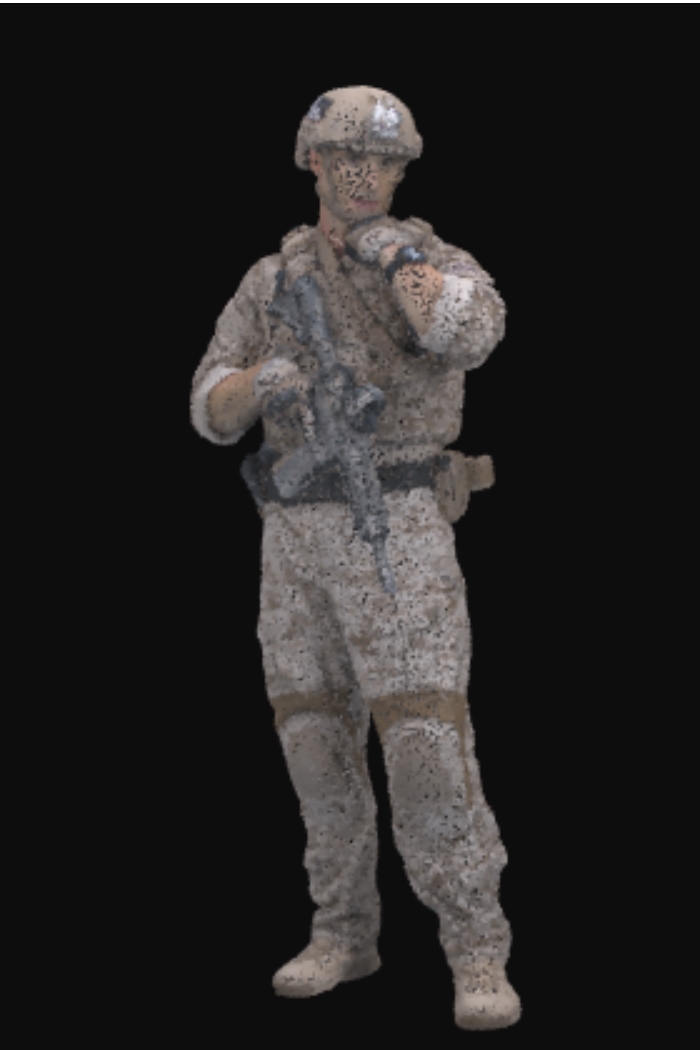}
    1M \hspace{2cm} 150K \hspace{2cm} 30K
         \vspace{-.2cm}
     \caption{Example visual view of a point cloud model with various densities.}
    \label{fig:subsample}
     \vspace{-.5cm}
\end{figure}

In order to decrease the density of point clouds, we propose three different approaches for spatial sub-sampling of dynamic point clouds. Our sub-sampling methods are based upon the concept of spatial clustering of neighbor points in the 3D space and sampling points in each cluster. 

\subsubsection{Algorithm 1}
Algorithm \ref{alg1} describes the process of our first sub-sampling approach. Each cluster consists of a set of $R$ neighbor points in the spatial domain with $R$ being the sub-sampling ratio. Clusters are simply constructed through points geometrically being sorted in the 3D space, firstly based on X component, then Y component, and finally based on Z component, therefore allowing each cluster with near-minimum spatial euclidean distance of consisting points. The process then simply selects a representative point from each cluster. Figure \ref{fig:subsample} demonstrates a visual view of how our sub-sampling approach works in practice. Figure \ref{fig:subsample} (left) shows a point cloud model consisting of 1.06 million points (the original model), while Figure \ref{fig:subsample} (middle) and Figure \ref{fig:subsample} (right) show sub-sampled models consisting of almost 150K and 30K points, corresponding to sub-sampling ratios of 7 and 35, respectively. While quality degradation on Figure \ref{fig:subsample} (right) can be noticed, there is likely not a noticeable visual difference between Figure \ref{fig:subsample} (left) and Figure \ref{fig:subsample} (middle) on this scale.

\subsubsection{Algorithm 2}
In our second sub-sampling methodology, we use the notion of histogram, and extend it to the context of point clouds to design a \textit{low-resolution density tree} to enhance the spatial clustering process. Our approach first computes the bounding box of all points in the point cloud model through determining the minimum and maximum point coordinates. We then divide the used space of the bounding box into a 3D voxel grid with a specific resolution, and create a density histogram. We process all points to determine their cell position and increment the number of points in the cell. With this, we create a low resolution density histogram, with higher number of points in a cell representing a denser cluster. We continue the process and recursively split each cell until its point count is lesser than a given threshold $m$. The choice of $m$ depends on the usage and the distribution of points in the point cloud model. Our clustering data structure is similar to an octree design given its recursive spatial division nature, except that it's a low-resolution octree built only for the purpose of clustering, in which each node contains a bounded number of points. If $m=1$, then the process would generate a complete octree.

\begin{figure}[!t]
     \vspace{-.5cm}
    \centering
    \includegraphics[width=.9\columnwidth]{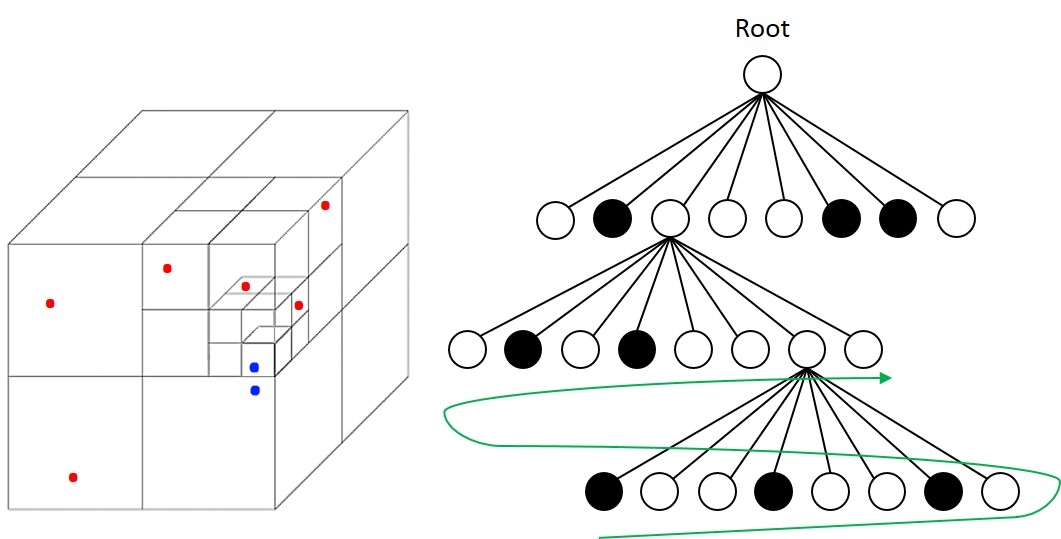}
         \vspace{-.2cm}
     \caption{Example illustration of sub-sampling within a density tree.}
    \label{fig:cubetree}
     \vspace{-.5cm}
\end{figure}

After the construction of the low-resolution density histogram, we perform a depth-first cluster sub-sampling process starting from the deepest layer of the tree. Starting from the left-most leaf, our process is to pick $k=\lceil{m/R}\rceil$ number of points in the cluster, given $R$ as the sub-sampling ratio. To select the points, we perform the sub-sampling process in Algorithm 1, sort the points in the 3D space, and uniformly pick representative points in the sorted list and removes the rest. The process then iterates through the next tree leaf, and continues towards the last tree node in the same depth level and then higher layers until exactly $\frac{(R-1)\times n}{R}$ points have been removed. The constructed tree leaves store the points of the point cloud model, with each leaf consisting of at most $m$ number of points. If $m=1$, a complete octree is generated, and our process removes the leaves from the deepest layers until the removal budget is satisfied. Figure \ref{fig:cubetree} describes a visual illustration of our process.

\subsubsection{Algorithm 3}
For further improvement of our sub-sampling approach, we enhance the clustering approach in Algorithm 2 and propose a more complex octree-based sub-sampling algorithm. To illustrate an example, in Figure \ref{fig:cubetree}, while the two blue points are closest neighbors, they can be ignored as a cluster when an octree or our low-resolution density tree is used. To better enhance our clustering algorithm, we follow the recursive steps proposed in Algorithm 2, and set $m=1$ to construct a complete octree. Our octree leaves store points of the point cloud model, with each point stored in one and only one leaf, and each leaf storing at most one point. Starting from the deepest left-most leaf, our approach calculates a list of nearest neighbors for each leaf, so that for a cluster of size $m$, we find and store $m-1$ nearest neighbors for each leaf point $p_i$. As a part of the process of the nearest neighbor search, our approach finds a closest leaf $p'_i$ corresponding to $p_i$ and marks it as processed. Once a cluster is constructed, we follow the sub-sampling process in Algorithm 1, and sort the points in the 3D space. We then pick the middle point as a representative points in the sorted list and remove the remainder. Similar to Figure \ref{fig:cubetree}, the process continues to the next octree leaf in the same depth followed by higher layers until exactly $\frac{(R-1)\times n}{R}$ points have been removed, given $R$ as the sub-sampling ratio.


Our sub-sampling approaches are all deterministic, with the output point cloud carrying exactly the determined percentage of the point volume. They are light-weight, and are developed in native environment with user API which are especially useful in large-scale dynamic point cloud sub-sampling. It should be noted that our adaptations are general, and are independent of the underlying sub-sampling methodology. Therefore, any sub-sampling approach can be employed.

\subsection{Human Visual Acuity}
\label{acuity}

\begin{figure}[!t]
    \centering
    \includegraphics[width=.6\columnwidth]{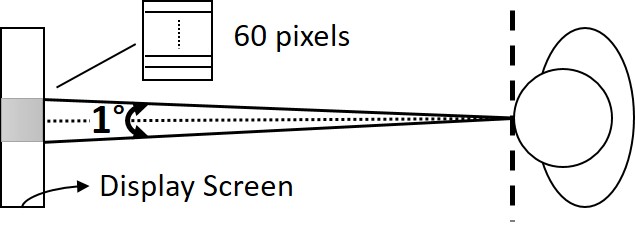}
     \caption{Human visual acuity.}
    \label{fig:acuity}
     \vspace{-.3cm}
\end{figure}

\begin{figure}[!t]
    \centering
    \includegraphics[width=.32\columnwidth]{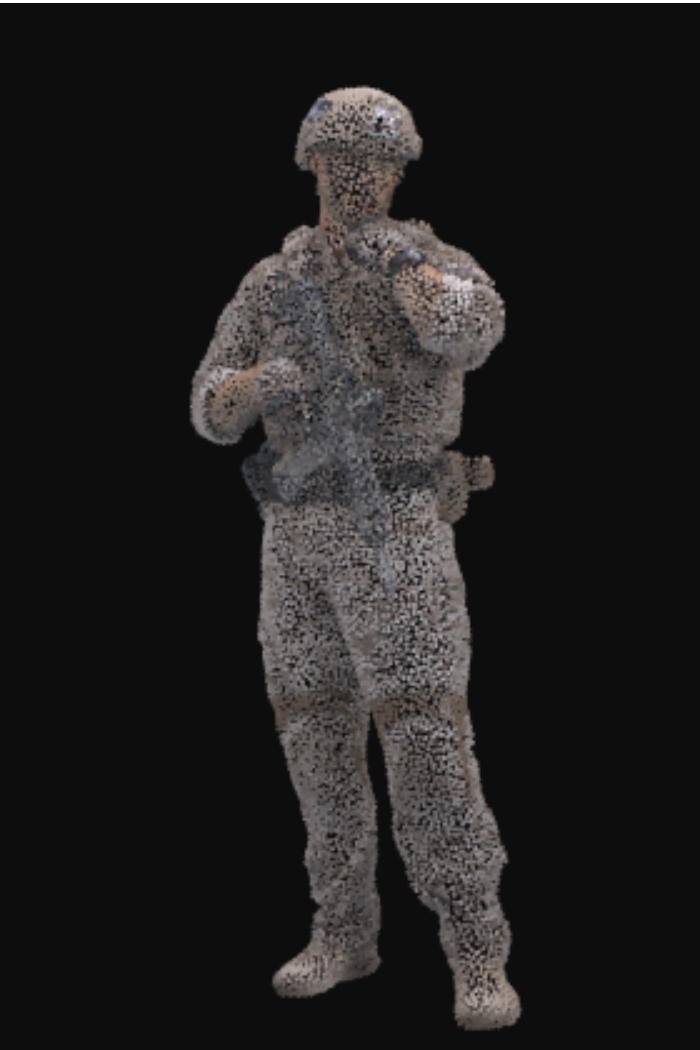}
    \includegraphics[width=.32\columnwidth]{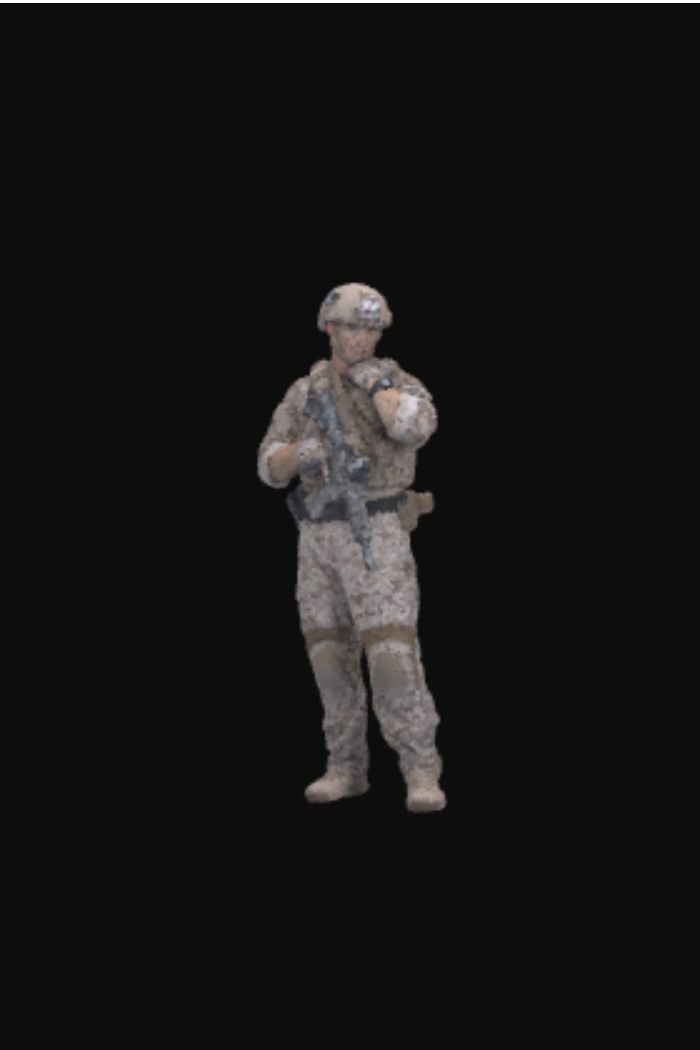}
    \includegraphics[width=.32\columnwidth]{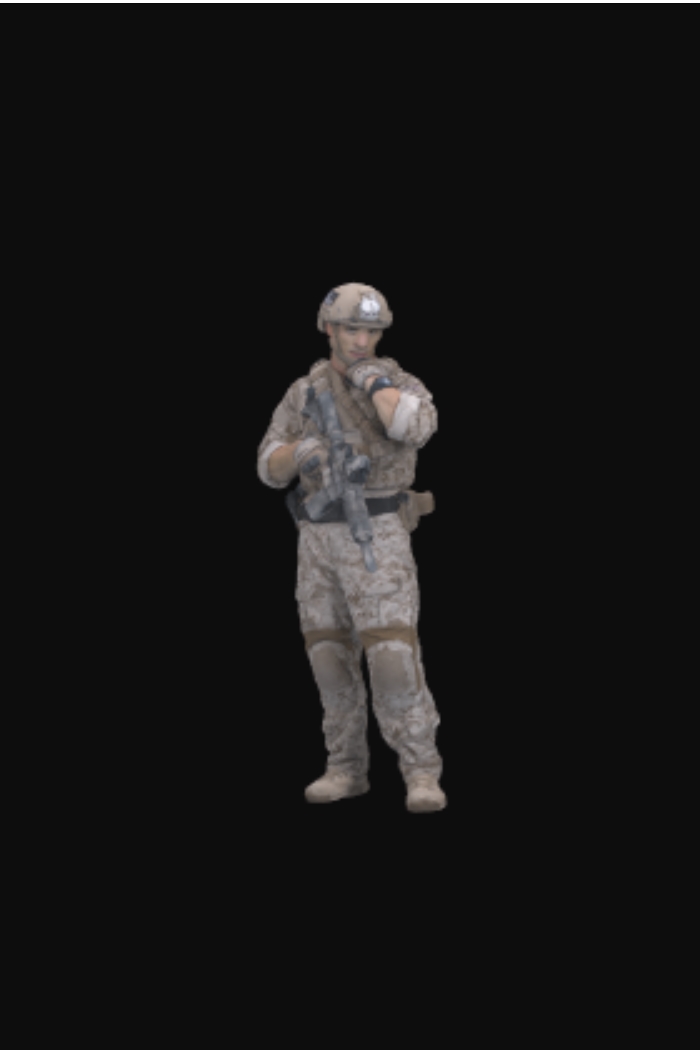}
         \vspace{-.2cm}
     \caption{Visual view of scaled point cloud models. (left) 15K sub-sampled, (middle) 15K scaled by a ratio of $2x$, (right) 1M scaled by a ratio of $2x$.}
    \label{fig:subsample2}
     \vspace{-.5cm}
\end{figure}

A point cloud density characteristic can be especially useful in the context of human visual acuity. Based on the features of human visual system, the perceived visual density of a screen and hense, the amount of anti-aliasing possibly required to make any computer graphic objects visually look convincing and smooth for users, depends on the pixel density of the screen and the object's distance from the user's eyes. While visual acuity varies individually and changes over time, the normal visual acuity for adults is 1 arc-minute in size, or 60 pixels per degree \cite{acuity}. Figure \ref{fig:acuity} illustrates the concept. As such, given $D$ as the distance of a user from the display screen, we can calculate the minimum Pixel per Inch (PPI) by:

\begin{equation}
PPI = \frac{1}{2\times D \times tan(\frac{1}{2}\times \frac{1}{60}\times \frac{\Pi}{180})}
\label{acuity1}
\end{equation}

In the context of point clouds in a 3D space, a point cloud object can scale up or down, or hold a distance from the viewport camera. Given $S$ as the scale factor of a point cloud object and $D'$ as the distance of the camera position from the centroid of the object's bounding box, we extend Eq. \ref{acuity1} as the following:

\begin{equation}
PPI = \frac{S}{2\times (D+D') \times tan(\frac{1}{2}\times \frac{1}{60}\times \frac{\Pi}{180})}
\label{acuity2}
\end{equation}

We use Eq. \ref{acuity2} to link visual acuity to a point cloud density level, which can serve as the basis for our optimization and density adaptations depending on the distance of the user from the screen, distance of the object from the viewport camera, as well as the scale of the model. Based on this finding, the use of a point cloud model with an average PPI of more than 60 is likely not visually noticeable for a user, and therefore can exhaust valuable limited resources. Similarly, if the total distance of the user from an object doubles and the point cloud object scales by a factor of 2, the same visual perception is maintained.

Figure \ref{fig:subsample2} compares the visual view a point cloud sub-sampled with a high ratio of 70 (15K points) in scaled and non-scaled views, compared to a scaled version of the non-scaled point cloud (1M points). While the quality degradation in Figure \ref{fig:subsample2} (left) is poor, Figure \ref{fig:subsample2} (middle) and Figure \ref{fig:subsample2} (right) are likely indistinguishable in such scale.

\subsection{User API}
For further convenience of generating various level-of-density representations for dynamic sequences as well as static point clouds, we have developed a stand-alone executable tool in Java 1.8.0, and also designed a set of simple APIs for users to construct desirable representations based on sub-sampling and scaling, which includes the following APIs:

\begin{itemize}
\item \texttt{subsample(int percentage, String inputpc, bool isIterative)}: generates a sub-sampled point cloud model with the percentage integer specifying the density of the output point cloud model relative to the input point cloud. The \texttt{isIterative} flag specifies if the process is iterated over multiple successive frames.
\item \texttt{scale(int percentage, String inputpc, bool isIt-\\erative)}: generates a scaled point cloud model given the scaling ratio.
\item \texttt{optimize(int distance, int scale, int inputpc, bool isIterative)}: automatically generates a sub-sampled point cloud with optimum PPI density given the distance (in inch) and/or the scale.
\end{itemize}

\section{Evaluation}
For evaluation, we used point cloud Library (PCL 1.8.1) to develop a dynamic point cloud visualization and streaming prototype and run our benchmarks. For the benchmarks, we used four different sequences provided by JPEG Pleno Database containing 8i voxelized dynamic point cloud sequences known as longdress, loot, redandblack, and soldier. In each sequence, the full object is captured by 42 RGB cameras configured at 30 fps, over a 10 second period. One spatial resolution in a cube of 1024x1024x1024 XYZRGB-formatted voxels is provided for each sequence. Table \ref{sequences} provides detailed information about our test sequences. We ran our experiments on an Intel Xeon E5520 64-bit machine with quad-core 2.27 GHz CPU, 12 GB RAM, and Gallium 0.4 on AMD GPU running Ubuntu 16.04 LTS, and used our defined manifest to describe various quality representations.

To focus on the amount of bandwidth savings, rendering performance, and objective quality measurement for this pilot study, we used our machine as the local HTTP streaming server to filter the negative impacts of network latency and bandwidth variations on the experimental results.

\begin{table}[!t]

\caption{Our point cloud test sequences}
\label{sequences}
\centering
\begin{tabular}{lll}
\hline
Test Sequence 	& Density 		& Average Volume (KByte/frame) \\
\hline
Red\& Black		& 700,531     & 14,731   \\
Loot 			& 778,467     & 16,515   \\
Soldier			& 1,060,464   & 22,937   \\
Longdress       & 794,641     & 17,456   \\
\hline
\end{tabular}
\vspace{-.3cm}
\end{table}

\begin{figure}[!t]
\centering
\includegraphics[width=\columnwidth, trim = 50 273 50 273, clip = true]{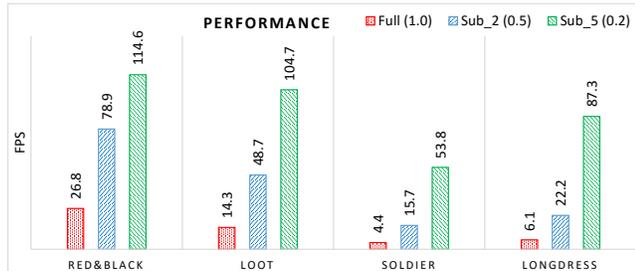}
\caption{A comparison of point cloud rendering performance in terms of FPS under 3 different sample representations.}
\label{results}
\vspace{-.5cm}
\end{figure}

\begin{figure}[!t]
\centering
\includegraphics[width=\columnwidth, trim = 50 273 50 273, clip = true]{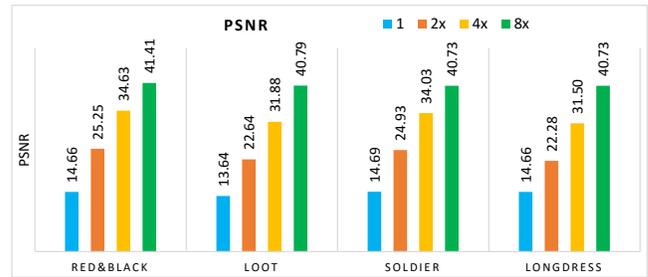}
\caption{A comparison of geometry PSNR under different scaling variances.}
\label{psnr}
\vspace{-.5cm}
\end{figure}

For the purpose of quantitative analysis, we used our developed tool to apply four sub-sampling ratios to the test sequences, and prepared 5 different representations in total for each point cloud frame ($REP_1$ to $REP_5$ representing highest to lowest density, given sub-sampling ratios of 1 to 5). The main goal was to study how our adaptations affect the total bandwidth and the rendering performance as well as the perceived quality. We collected statistics for the total bandwidth usage, rendering performance, and quality measurements, and compared our results with the baseline scenario where no adaptation is employed.

Each trial of our experiment was run for a total of 30 consecutive test sequences, and we repeated each experiment 6 times to ensure that the standard deviations for variable metrics are within acceptable limits. As obviously the total bandwidth saving is directly correlated to the sub-sampling ratio, a total bandwidth saving of approximately \%50 and \%80 was achieved in raw point cloud frames given sub-sampling ratios of 2 and 5, respectively. We measured the client's rendering performance in terms of the average FPS continuously for 100 times during each session, and recorded the data when adaptations are applied (lower representations $REP_2$ and $REP_5$ with sub-sample ratios of 2 and 5) compared to the baseline case where no adaptation is applied ($REP_1$). Figure \ref{results} demonstrates results for only a subset of our experiments on all of our benchmarks under Algorithm 1. The results show that depending on the scene, our adaptations can significantly improve the processed FPS up to more than 10x compared to the baseline case where no adaptation is applied.

To verify our assumptions for visual acuity, we ran extensive measurements on the objective quality using PSNR of point-to-point distortions within our adaptations. We used the official MPEG PCC Quality Metric Software \cite{mpegpcc}, which compares an original point cloud with an adapted model and provides numerical values for point cloud PSNR. We averaged the PSNR measurements over all dynamic point cloud frames, and collected the experimental results. Figure \ref{psnr} illustrates how the PSNR of our adaptations (an original $REP_1$ point cloud model against a $REP_2$ sub-sampled model carrying \%50 of the original points) is affected under different scaling variants, for 2x, 4x, and 8x scale-down ratios. The results demonstrate interesting insights on the trade-offs of a point cloud quality, scaling, and sub-sampling ratios, and confirms our earlier assumptions in regards to the effects of scaling point clouds on the human's visual acuity.

\begin{figure}[!t]
\centering
\includegraphics[width=.32\columnwidth]{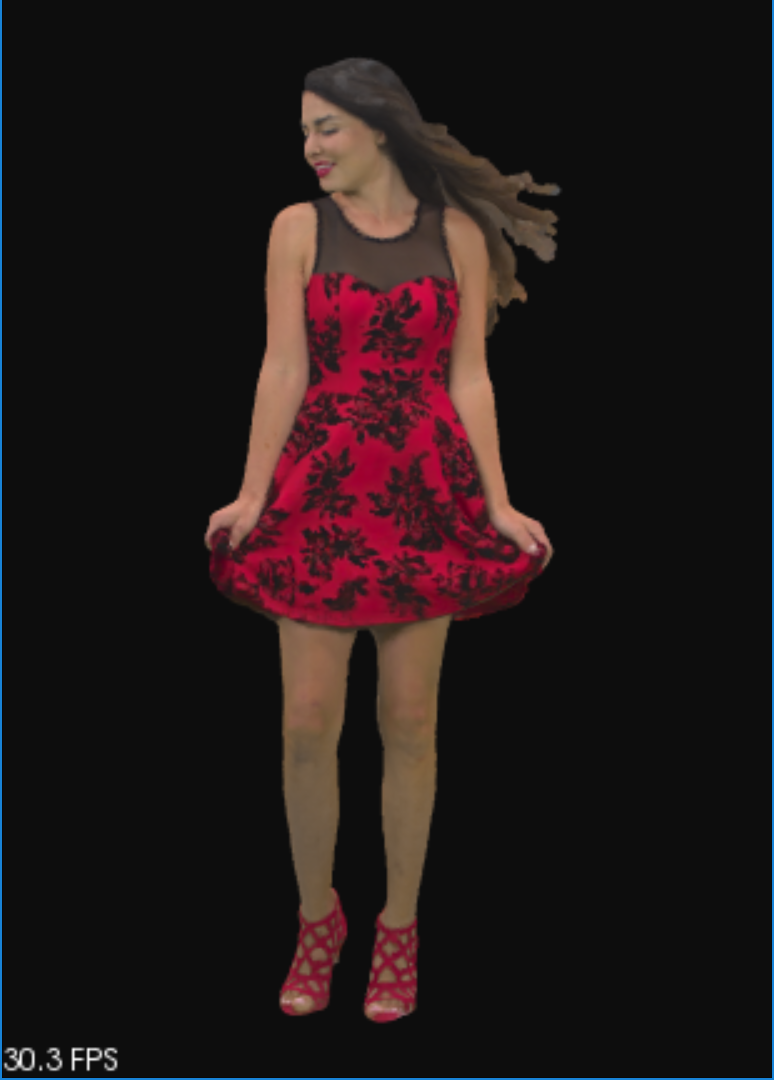}~\includegraphics[width=.32\columnwidth]{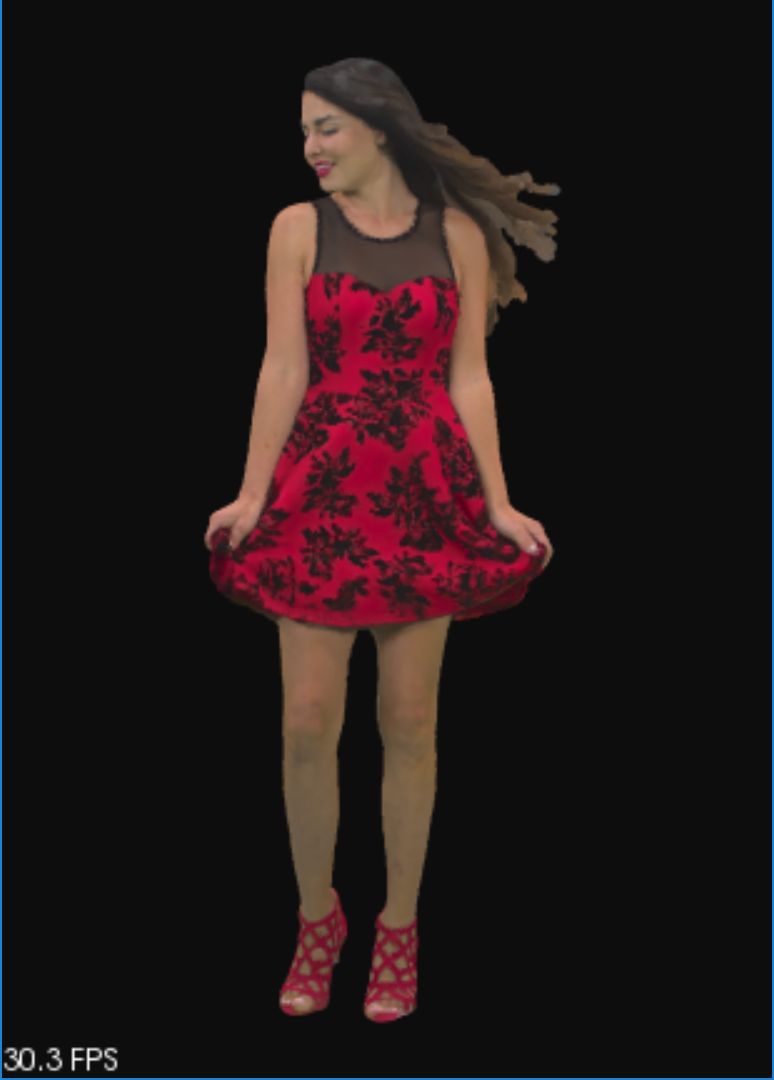}~\includegraphics[width=.32\columnwidth]{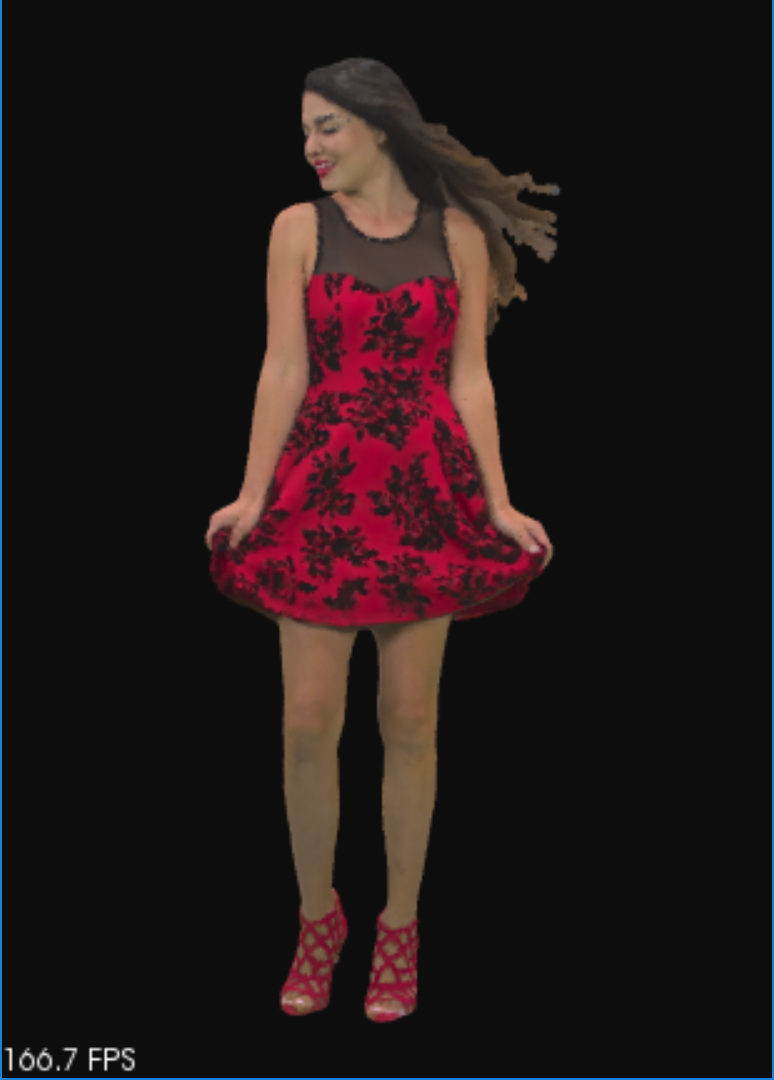}
\caption{Visual comparison of a specific frame within \textit{Red\& Black} under 3 sample representations: (left) $REP_1$, (middle) $REP_2$, (right) $REP_5$.}
\label{redblack}
\vspace{-.3cm}
\end{figure}

Figure \ref{redblack} demonstrates a sample screenshot of our experiments on \textit{Red\& Black} illustrating 3 consecutive frames from highest representation ($REP_1$) (left) to the lowest representation ($REP_5$) (right). As can be seen, depending on a point cloud distance and scale, our adaptation can result in negligible noticeable quality impacts, sometimes not even perceptible, while it maintains possibly highest quality to ensure a satisfactory user experience. Furthermore, given the reduction in the rendering overhead, our adaptation makes it possible to more efficiently stream and render multiple dense point cloud objects within a single viewing scene, which previously was not possible due to the limited hardware resources for streaming and processing unnecessary bulky point clouds. Overall, considering the significant bandwidth saving, performance improvement, and lower hardware exhaustion achieved using our adaptations, it is reasonable to believe that many would accept our adaptations to respect the valuable limited resources in an immersive environment.

\section{Conclusion and Future Work}
In this pilot study, we proposed dynamic point cloud streaming adaptation techniques to tackle the high bandwidth and processing requirements of transmission and rendering of dense point clouds. Our novel adaptations exploits the semantic link of adaptive streaming with a user's visual acuity, limited bandwidth, and other constrained resources to provide dynamic adaptations in the context of point cloud streaming. Our initial experimental results show that our adaptations can significantly save point cloud streaming bandwidth and improve rendering performance with negligible noticeable quality impacts.

We are currently working on different rate adaptation algorithms and study their various quality trade-offs.

\begin{acks}
This work was supported in part by the Austrian Research Promotion Agency (FFG) under the Next Generation Video Streaming project "PROMETHEUS".
\end{acks}

\balance
\bibliographystyle{ACM-Reference-Format}
\bibliography{ref}

\end{document}